\documentclass[aip, apl, preprint]{revtex4-1}

\usepackage{graphicx}
\usepackage{txfonts}
\usepackage{color}

\draft % marks overfull lines with a black rule on the right

\begin{document}

%\preprint{}

\title{\textcolor{blue}{Anomalous dip observed in intensity autocorrelation function
as an inherent nature of single-photon emitters}}

\author{H. Nakajima}
\email{nakajima@es.hokudai.ac.jp}
\affiliation{Research Institute for Electronic Science, Hokkaido University, Sapporo 001-0021, Japan}
\affiliation{Research Fellow of the Japan Society for the Promotion of Science, Tokyo 102-8472, Japan}
\author{H. Kumano}
\affiliation{Research Institute for Electronic Science, Hokkaido University, Sapporo 001-0021, Japan}
\author{H. Iijima}
\affiliation{Research Institute for Electronic Science, Hokkaido University, Sapporo 001-0021, Japan}
\author{I. Suemune}
\affiliation{Research Institute for Electronic Science, Hokkaido University, Sapporo 001-0021, Japan}
%\thanks{}


\date{\today}

\begin{abstract}
We report the observation of an anomalous antibunching dip in intensity autocorrelation function
with photon correlation measurements on a single-photon emitter (SPE). 
We show that the anomalous dip observed is a manifestation of quantum nature of SPEs.
Taking population dynamics in a quantum two-level system into account correctly,
we redefine intensity autocorrelation function.
This is of primary importance for precisely evaluating the lowest-level probability
of multiphoton generation in SPEs
toward realizing versatile pure SPEs for quantum information and communication.
\end{abstract}

%\pacs{78.66.Fd, 85.60.Jb, 74.25.Jb}% insert suggested PACS numbers in braces on next line

\maketitle %\maketitle must follow title, authors, abstract and \pacs

A variety of single-photon emitters (SPEs)~\cite{Michler01, Zwiller02, Takemoto03, Claudon04, Pelton06, Intallura, Nothaft09, Kuhn10, Darquie, Keller11, Han13, Hausmann14, Beveratos, Ester, Rivoire22, Lochmann23, Becher30, Kumano28, Nakajima29, Lounis08}
have been widely investigated for applications in quantum key distribution (QKD)~\cite{Gisin15}, 
quantum information processing~\cite{Nielsen16}, and quantum metrology~\cite{Giovannetti17}.
Single-photon emission has been demonstrated by using quantum two-level systems formed
in single molecules~\cite{Lounis08, Nothaft09}, atoms~\cite{Kuhn10, Darquie},
ions~\cite{Keller11}, color centers in diamond~\cite{Han13, Hausmann14, Beveratos},
and semiconductor quantum dots (QDs)~\cite{Michler01, Zwiller02, Takemoto03, Claudon04, Pelton06, Intallura, Ester, Rivoire22, Lochmann23, Becher30, Kumano28, Nakajima29}.
Generating single-photon pure state is crucial for assuring the firm security
in the cryptography~\cite{Waks20}
and also minimizing error rate in linear optical quantum computing~\cite{Knill}.
Therefore, suppression of the multiphoton generation is strongly required for the practical SPEs.
Recently, with a variety of quantum systems, SPEs with considerably
low multiphoton probability have been reported~\cite{Darquie, Ester, Claudon04},
and implementation to the prototype QKD systems has also been demonstrated~\cite{Takemoto03, Beveratos, Intallura}.

Photons generated from SPEs are generally inspected
with the Hanbury-Brown and Twiss (HBT) setup~\cite{Brown18},
where photons separated into two arms are introduced to single-photon detectors
located on each arm for photon correlation measurements.
The intensity autocorrelation function~\cite{Glauber} is composed of
coincidence counts as a function of the delay time $\tau$
between photon detection events in each detector.
The coincidence counts at $\tau = 0$ exhibit a simultaneous photon detection by the two detectors.
Therefore, multiphoton generation can be directly measured with coincidence counts at $\tau=0$
~\cite{Michler01, Zwiller02, Pelton06, Lounis08, Han13, Hausmann14}
and this usually appears as a \textit{peak} in the intensity autocorrelation function.
%
%Needless to say, a \textit{peak}-shaped coincidence at $\tau\sim 0$
%is an essential prerequisite in evaluating the multiphoton contribution of SPEs.

In this paper, observation of counterintuitive \textit{dip}-shaped structure at $\tau\sim 0$
in intensity autocorrelation function is reported for the first time.
We show the dip structure originates from an inherent nature of a single quantum emitter.
In order to explicitly include population dynamics in a quantum two-level system,
we derive  an extended form of the conventionally used intensity autocorrelation function.
This provides a way to precisely determine the probability of generating single-photon pure states
from SPEs over a wide range of operating conditions.

\begin{figure}
\begin{center}
\includegraphics[bb=5 0 460 580, clip, width=8.8cm]{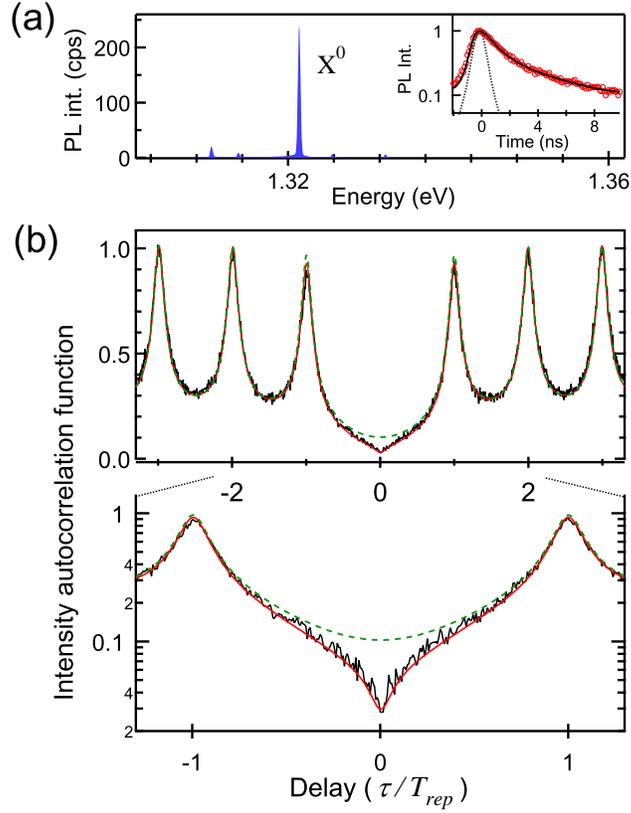}
\end{center}
\caption{
(a) $\mu$-PL spectrum from a single InAs QD.
Inset indicates a decay profile of the ${\rm X}^0$ emission line (red circles)
and fitted result (black line)
convoluted with a response function of our whole system (dotted line).
(b) Measured intensity autocorrelation function (black line)
for the ${\rm X}^0$ emission line with a time bin of 100 ps.
Simulated curves based on Eq.~(\ref{oldeq}) with $\alpha_0=0$ (green dashed line)
and Eq.~(\ref{neweq}) (red line) are also shown.
Both curves are convolved with a system response function.
Measured autocorrelation function has a cusp at $\tau = 0$, which is not consistent with Eq.~(\ref{oldeq}).
The red curve indicates the best fit to the measured function, which gives $\alpha_0=0.003$.
Expanded view at $\tau \sim 0$ in a logarithmic scale is displayed at the bottom.
}
\label{decay}
\end{figure}

InAs QDs grown on (001) GaAs by metalorganic molecular-beam epitaxy was used to realize a SPE.
For isolating a single QD, pillar structures with the diameter of 500 nm were formed
with reactive ion etching and were embedded with metal to enhance photon extraction efficiency.
Further details on sample preparation are given in refs.~\onlinecite{Kumano28} and ~\onlinecite{Nakajima29}.
Optical properties of the QDs were examined by a standard micro-photoluminescence ($\mu$-PL)
 setup equipped with a mode-locked Ti:sapphire laser
(photon energy of 1.3920 eV, pulse repetition period of 13.2 ns, pulse duration of $\sim$ 5 ps)
and a Si charge-coupled-device detector.
Figure \ref{decay} (a) shows a $\mu$-PL spectrum observed from a single QD at 20 K.
The excitation power was 2.1 $\mu$W which corresponds to
the average number of excitons ($\bar N_{\rm X}$) photoinjected into the QD of $\sim$ 0.2.
The emission line centered at 1.3214 eV is prominent and we focus on this line hereafter.
From the linear excitation power dependence of the PL intensity and
the presence of finite exciton fine structure splitting~\cite{Bayer},
this emission line was assigned to be a neutral exciton (${\rm X}^0$).

Under the same excitation condition, a photon correlation measurement
was carried out with the HBT setup employing a pair of single-photon counting modules (SPCMs).
Resultant intensity autocorrelation function is displayed as black line in Fig. 1 (b)
with its expanded view around zero delay in the lower trace.
The accumulation time for building up the histogram with a multi-channel scaler was about 10 h.
Strongly suppressed coincidence counts at $\tau \sim 0$ manifest highly pure single-photon emission
from the present SPE.

Here, we analyze the measured intensity autocorrelation function
with a commonly accepted formula
under nonresonant pulsed excitation~\cite{Han13,Beveratos, Nakajima29, Zwiller02, Rivoire22}
\begin{equation}
N^{ - 1}\left \{ B + {\alpha _0}\exp \left( - \frac {\left| \tau  \right|}{ \tau _{e}} \right )+\sum\limits_{n \ne 0} \alpha _n\exp \left ( - \frac {\left | \tau  - n \cdot {T_{rep}} \right |}{\tau _{e}} \right )   \right\},
\label{oldeq}
\end{equation}
where $\alpha_0$, $\alpha_{n(\neq0)}$, $T_{rep}$, $\tau _e$ and $N$ are
the degree of multiphoton contribution ($0\leq\alpha_0\leq1$),
correlation peak height of $n$-th excitation cycle
($\alpha_{n(\neq0)} \equiv 1$), repetition period of the excitation pulses, decay time constant of the emitter,
and the normalization factor, respectively.

Here, $B$ is the baseline originating from an accidental coincidence,
estimated to be $\sim$0.009~~\cite{comment2}.
As for $\tau_{e}$, we have independently measured the decay profile
of the ${\rm X}^0$ emission line (inset of Fig. 1 (a)),
and obtained double-exponential decay times of 0.9 and 6.1 ns.
The shorter decay component is the exciton lifetime commonly observed in InAs QDs~\cite{Becher30},
while the longer one is most probably due to additional transitions
involving other excitonic states, such as dark excitons~\cite{Dalgarno26} or charged excitons~\cite{Piechal27}.
Intensity autocorrelation function based on Eq.~(\ref{oldeq}) is simulated~\cite{comment4}
and the convoluted result with a system response function
is displayed as the green dashed line in Fig. 1 (b).
In this simulation, $\alpha_0$ was set to zero assuming an ideal SPE.
The overall properties are well reproduced.
However, the important finding is that the measured coincidence counts at $\tau \sim 0$
are lower than the one calculated with Eq.~(\ref{oldeq}) for the ideal SPE,
which yields an anomalous dip rather than usually observed peak at $\tau = 0$.
This is, to the best of our knowledge, the first observation of the anomalous dip structure
in the intensity autocorrelation function for SPEs.

\begin{figure}
\begin{center}
\includegraphics[bb=0 0 490 380, clip, width=8cm]{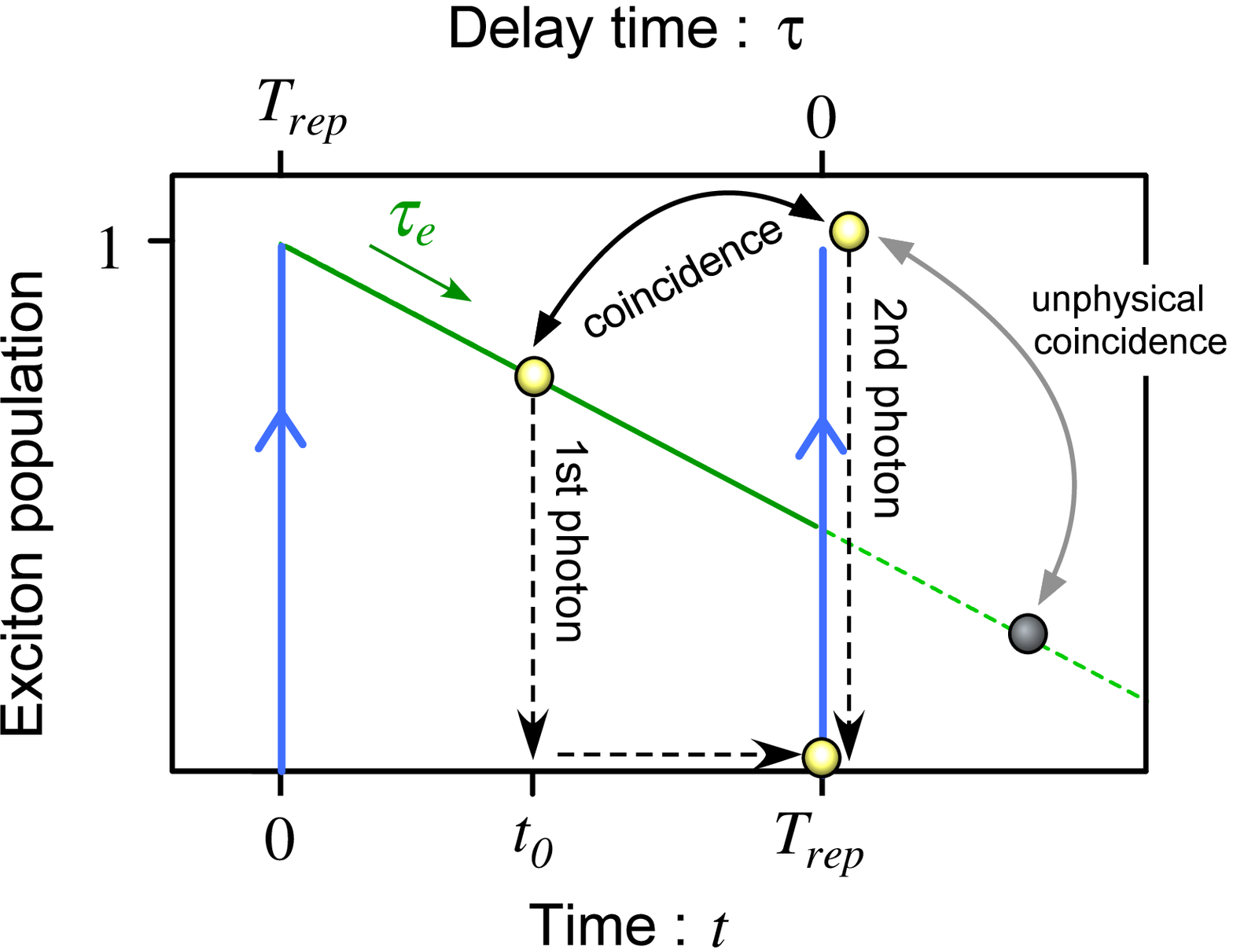}
\end{center}
\caption{
Schematic of the dynamics of exciton population in a QD.
Green thin line indicates the decay profile of exciton populated at $t=0$
as a function of $t$ (bottom axis) and delay time $\tau$ with respect to $t=T_{rep}$ (upper axis).
After the first photon emission at $t_0$,
system stays in the ground state until next excitation (black dashed arrow).
Black arrow indicates possible coincidence between the first and second photon emissions,
while the gray arrow corresponds to unphysical coincidence
in which the exciton decay is independent of the subsequent excitations as reflected in Eq. (1).
}
\label{autocorr}
\end{figure}

The observed dip-shaped coincidence with a cusp at $\tau = 0$ reveals that
there exists qualitative difference between the measured intensity autocorrelation function
and Eq.~(\ref{oldeq}).
We discuss, to clarify the difference,
the coincidence counts between photons labeled as the first and second photons
triggered by the different excitation pulses.
Assuming for simplicity that excitation pulse drives
the exciton population in a QD, $p_{|{\rm X}>}$, to unity
and the second photons are emitted instantaneously after excitation at $t=T_{rep}$
(Fig. 2).
The $p_{|{\rm X}>}$ initiated to unity at $t=0$
will relax to the ground state (GS) with decay time constant of
$\tau_e$ as indicated by the green line in Fig. 2.
%This process is described by a simple rate equation of a two-level system
%$dp_{|{\rm X}>}(t)/dt=-p_{|{\rm X}>}(t)/\tau_e$,
%and the decay of population probability is given by $\exp(-t/\tau_e)$
%
The coincidence between the first and second photons
indicated by the black arrow occurs at the delay time of $\tau=T_{rep}-t_0$,
and its counts are proportional to $\exp(-t_0/\tau_e)$,
which is the exciton population at the time of the first photon emission.
Provided that the exciton populated at $t=0$ decays independently of the next excitation
at $t=T_{rep}$,
the coincidence counts could be recorded even for $t \ge T_{rep}$ ($\tau \le 0$)
as shown in the gray arrow.
The conventional formula Eq.~(\ref{oldeq}) is formulated under this situation,
in which all the contribution of the photon pairs to the coincidence counts
is summed up uncorrelatedly to the population dynamics
which takes place in the excitation and emission processes
(green dashed curve in Fig. 1(b)).
In a realistic quantum two-level system, in contrast, 
once the first photon is emitted at the time $t_0$ ($0 \leq t_0 \leq T_{rep}$),
the population in a QD is reset to the GS and keeps in the GS
until experiencing the next excitation at $t=T_{rep}$ as displayed by the black dashed 
arrow in Fig. 2.
Therefore, no coincidence count is possible for $\tau \le 0$.
This gives rise to the essential difference between
observed intensity autocorrelation function
and Eq. ~(\ref{oldeq}).

The population decay for $\tau \le 0$ which brings unphysical coincidence counts
is expressed by $\exp(-|\tau|/\tau_e)$ (green dashed line)
with its amplitude being normalized by the exciton population at $\tau = 0$.
Since the sum of population probability over two states in the quantum two-level system
is unity for arbitrary delay time of $\tau$,
physically valid coincidence is given by
the complementary counterpart of the unphysical coincidence,
i.e., $1 \!- \! \exp(-|\tau|/\tau_e)$.
This term corresponds to the modulation intensity
to apply to the unphysical coincidence counts
given at each delay time of $\tau$ in Eq.~(\ref{oldeq}),
so that the quantum nature of the emitter as a two-level system
is appropriately incorporated into the intensity autocorrelation function.
This is the brief interpretation to the observed anomalous dip structure with a cusp at zero delay.
%
%Eq. 1 is composed of uncorrelated photon pairs originated from different excitation pulses
%in that the population generatd by an excitation event decays independently of
%the subsequent excitation processes.

According to the above argument,
we define an extended intensity autocorrelation function
including the population dynamics in a quantum two-level system as
\begin{equation}
\tilde g^{(2)}(\tau \ge 0) =N^{ - 1}\left \{  B + {\alpha _0}\exp \left( - \frac {\left| \tau  \right|}{ \tau _{e}} \right )+\sum\limits_{n > 0} \alpha _n\exp \left ( - \frac {\left | \tau  - n \cdot {T_{rep}} \right |}{\tau _{e}} \right ) \cdot  [1-\exp(- |\tau| / \tau _e) ] \right \},
\label{neweq}
\end{equation}
and $\tilde g^{(2)}(\tau \leq 0) = \tilde g^{(2)}(-\tau)$~\cite{comment5}.
In comparison to Eq.~(\ref{oldeq}), the anomalous dip observed at $\tau \sim 0$ is
satisfactorily reproduced with the $\tilde g^{(2)}(\tau)$ as indicated by the red line in Fig. 1 (b).
Furthermore, the extended function allows us to precisely determine the
multiphoton contribution of $\alpha_0=0.003$ which cannot be derived with Eq.~(\ref{oldeq}).
These results demonstrate that considering the population dynamics,
as an inherent nature of quantum emitters,
is essential for evaluating the intensity autocorrelation function under the pulsed excitation.

In what follows,
we discuss the condition for emerging the anomalous antibunching dip based on Eq.~(\ref{neweq}).
The anomalous dip is caused by applying the modulation term $1\!-\!\exp(-\tau/\tau_e)$
to unphysical coincidence counts characterized by exciton population
at $t=T_{rep}$, i.e., $\exp(-T_{rep}/\tau_e)$ (see Fig. 2 and Eq.~(\ref{oldeq})).
Thus, for evaluating the $\alpha_0$, it is beneficial to describe the dip depth
as a function of $T_{rep}/\tau_e$ which is specified by selecting the emitter
and the repetition period of the excitation.
Here, we introduce the dip depth defined by $\Delta - \tilde g^{(2)}(0) = \Delta - \alpha_0$,
where $\Delta$ is the lower limit of the coincidence counts at $\tau = 0$
without considering the inherent nature of quantum emitter~\cite{comment3}, and we set $B = 0$.
Figure 3 presents the calculated dip depth as a function of $T_{rep}/\tau_e$
for some $\alpha_0$ values.
In this figure, all traces tend to $-\alpha_0$ for sufficiently high $T_{rep}/\tau_e$,
which indicates that peak-shaped coincidence with the amplitude of $\alpha_0$
appears as the multiphoton contribution.
In this condition, Eq.~(\ref{neweq}) reduces to Eq.~(\ref{oldeq}).
Actually, in most of reports, $\alpha_0$ has been evaluated
with relatively high $T_{rep}/\tau_{e}$ region
such as $>10$~\cite{Michler01, Zwiller02, Takemoto03, Lounis08, Keller11, Han13}.
However, for the low $T_{rep}/\tau_e$ region, dip-shaped coincidence emerges.
This is because the coincidence counts based on
the uncorrelated decay (green dashed line in Fig. 2) are overestimated,
and the amplitude of modulation required to include the quantum nature
is enhanced for the low $T_{rep}/\tau_e$.
Thus, the conventional formula Eq.~(\ref{oldeq}) is no longer valid.
In the present case, since $T_{rep}/\tau_{e} \sim 2.2$ and $\Delta > \alpha_0$,
the anomalous dip was clearly observed as indicated by Fig. 1 (b). 
Therefore, it is essential to employ the $\tilde g^{(2)}(\tau)$
especially for the SPEs with low $\alpha_0$ operating with low $T_{rep}/\tau_{e}$ conditions
such as high repetition cycles.~\cite{Rivoire22, Lochmann23}

\begin{figure}
\begin{center}
\includegraphics[bb=10 0 380 410, clip, width=7.6cm]{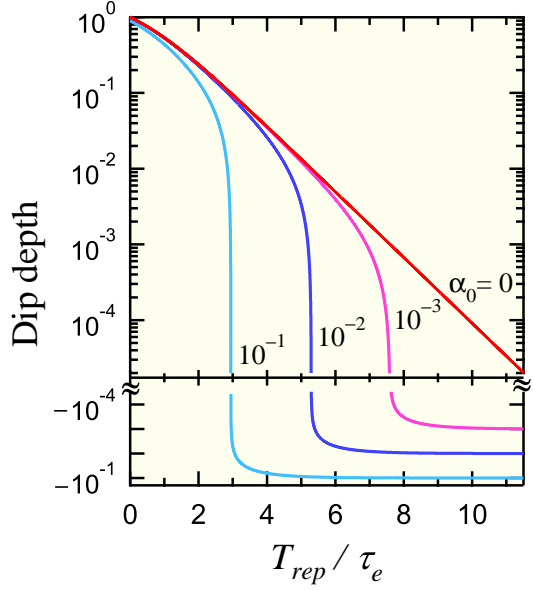}
\end{center}
\caption{
Calculated depth of anomalous dip at $\tau = 0$
as a function of $T_{rep}/\tau_e$ for specific $\alpha_0$ values.
In this calculation, $B=0$ is assumed and system response function is not taken into account.
Negative value represents peak-shaped coincidence at $\tau \sim 0$.
The dip structure will be distinct for SPEs with low $\alpha_0$
operating with low $T_{rep}/\tau_e$ conditions.
For sufficiently high $T_{rep}/\tau_e$, the dip depth approaches to $-\alpha_0$.
Generalized intensity autocorrelation function presented in this work is applicable to
the whole $T_{rep}/\tau_e$ range, which is essential to precisely evaluate the $\alpha_0$
under arbitrary operating conditions.
}
\label{setup}
\end{figure}

Note that the fine fitting for the height of each correlation peak at
$\tau =n \! \cdot \! T_{rep}$ $(|n| \ge 1)$
shown in Fig. 2 is due to relatively low excitation condition such that $\bar N_{\rm X} \sim 0.2$.
For larger excitation power, the peak heights are subject to the effect of excitation rate of $G$
as is the case with the well-known antibunching lineshape in a single-photon emission
under cw excitation.~\cite{Lounis}
On the other hand, the derived modulation term $1-\exp(-|\tau|/\tau_e)$
is irrelevant to the $G$ for $\tau \sim 0$ since the system is free from excitation.

In conclusion, we have reported the observation of an anomalous antibunching dip
in intensity autocorrelation function with a semiconductor single-photon emitter.
By redefining the autocorrelation function to include the population dynamics
in quantum emitters, the observed dip was clearly interpreted.
Applying the extended autocorrelation function
to the result of the photon correlation measurements enables us to successfully evaluate
one of the most important figure of merit $\alpha_{0}$ even with relatively
low $T_{rep}/\tau_{e}$ condition evoking a dip at around zero delay.
Our findings are invaluable to deal with versatile single-photon emitters
demanded for the state-of-the-art quantum information devices.

The authors would like to acknowledge Dr. H. Sasakura and  Dr. C. Hermannst\"{a}dter for fruitful discussions.
This work was supported in part by the Grant-in-Aid for Scientific Research (B), No.24310084, (S), No.24226007,
Hokkaido Innovation Through Nanotechnology Supports (HINTs) from the Ministry of Education, Culture, Sports,
Science and Technology, and Strategic Information and Communications R\&D Promotion Programme (SCOPE)
from Ministry of Internal Affairs and Communications.

%\bibliography{nakabibfile}

\begin{thebibliography}{99}
\bibitem{comment4}The emitter employed in this work shows double-exponential decay,
thus the term $\exp(-1/\tau_e)$ in Eq.~(\ref{oldeq}) 
can be rewritten as $A_1\exp(-1/\tau_1)+A_2\exp(-1/\tau_2)$,
where $A_1$, $A_2$, $\tau_1$, $\tau_2$ are 0.69, 0.31, 0.9 ns, 6.1 ns, respectively.
\bibitem{comment5} In the redifined intensity autocorrelation function,
symmetrization operation to be even function in terms of $\tau$ is carried out
after summing up all the contribution of physically possible coincidence counts
triggered by excitations for $n > 0$.
\bibitem{comment3}Function value of Eq.~(\ref{oldeq}) at $\tau=0$ corresponds to $\Delta$
with $\alpha_0=B=0$.
Using $\Sigma$ defined as
$ 2 \! \cdot \! \sum\limits_{n > 0} \alpha_n \exp(- n \! \cdot \! T_{rep}/\tau_e)$,
we have $\Delta=\Sigma/(1+\Sigma)$.
\end{thebibliography}

\end{document}